\documentclass[11pt]{elsart}
\usepackage{amsfonts}
\usepackage{slashbox}
\usepackage{graphicx}
\usepackage{amssymb,amsmath}% AMS packages
\usepackage{mathrsfs} % mathcal fonts

\begin{document}

\begin{frontmatter}
\title{Traditional sufficient conditions for Nash implementation may fail on Internet}
\author{Haoyang Wu\corauthref{cor}}
\corauth[cor]{Wan-Dou-Miao Research Lab, Suite 1002, 790 WuYi Road,
Shanghai, 200051, China.} \ead{yangki76@163.com} \ead{Tel:
86-18621753457}

\begin{abstract}
The Maskin's theorem is a fundamental work in the theory of
mechanism design. In this paper, we propose that if agents report
messages to the designer through channels (\emph{e.g.}, Internet),
agents can construct a self-enforcing agreement such that any
Pareto-inefficient social choice rule satisfying monotonicity and
no-veto will not be Nash implementable when an additional condition
is satisfied. The key points are: 1) The agreement is unobservable
to the designer, and the designer cannot prevent the agents from
constructing such agreement; 2) The agents act non-cooperatively,
and the Maskin mechanism remain unchanged from the designer's
perspective.
\end{abstract}
\end{frontmatter}

\section{Introduction}
The theory of mechanism design has been developed and applied to
many branches of economics for decades. Nash implementation is a
cornerstone of the mechanism design theory. The Maskin's theorem
\cite{Maskin1999} provides an almost complete characterization of
social choice rules (SCRs) that are Nash implementable: when the
number of agents is at least three, the sufficient conditions for
Nash implementation are monotonicity and no-veto, and the necessary
condition is monotonicity. Note that an SCR is specified by a
designer, a desired outcome from the designer's perspective may not
be desirable for the agents (See Table 1 in Section 3.1).

The Maskin mechanism (page 394, \cite{Serrano2004}) constructed in
the proof of Maskin's sufficiency theorem is an abstract mechanism.
People seldom consider how the designer actually receives messages
from agents. Roughly speaking, there are two distinct manners:
direct and indirect manner. In the former manner, agents report
their messages to the designer directly (\emph{e.g.}, speak face to
face, hand over, \emph{etc}), thereby the designer can know exactly
that a message is reported by an agent himself, not by any other
device. In the latter manner, agents report messages to the designer
through channels (\emph{e.g.}, Internet, cable \emph{etc}). Thereby,
when the designer receives a message from a channel, he cannot know
what has happened on the other side of the channel: whether the
message is reported by an agent himself, or generated by some device
authorized by an agent.

Traditionally, nobody notice the difference between the two manners
in the Maskin mechanism. However, in this paper, we will point out
that traditional sufficient conditions on Nash implementation may
fail if agents report messages to the designer in an indirect
manner. The rest of the paper is organized as follows: Section 2
recalls preliminaries of the mechanism design theory given by
Serrano \cite{Serrano2004}; Section 3 is the main part of this
paper, where we will propose a self-enforcing agreement to help
agents break through the restriction of Maskin's sufficiency
theorem. Section 4 draws the conclusion.

\section{Preliminaries}
Let $N=\{1,\cdots,n\}$ be a finite set of \emph{agents} with $n\geq
3$, $A=\{a^{1}, a^{2},\cdots\}$ be a finite set of social
\emph{outcomes}. The information held by the agents is summarized in
the concept of a \emph{state}. The true state is not verifiable by
the designer. We denote by $t$ a typical state and by $\mathcal{T}$
the domain of possible states. At state $t\in\mathcal {T}$, each
agent $j\in N$ is assumed to have a complete and transitive
\emph{preference relation} $\succeq_{j}^{t}$ over the set $A$. We
denote by $\succeq^{t}=(\succeq_{1}^{t},\cdots,\succeq_{n}^{t})$ the
profile of preferences in state $t$, and denote by $\succ_{j}^{t}$
the strict preference part of $\succeq_{j}^{t}$.

Fix a state $t$, we refer to the collection $E=\langle
N,A,(\succeq_{j}^{t})_{j\in N}\rangle$ as an \emph{environment}. Let
$\varepsilon$ be the class of possible environments. A \emph{social
choice rule} (SCR) $F$ is a mapping $F:\varepsilon\rightarrow
2^{A}\backslash\{\emptyset\}$. A \emph{mechanism}
$\Gamma=((M_{j})_{j\in N},g)$ describes a message or strategy set
$M_{j}$ for agent $j$, and an outcome function $g:\prod_{j\in
N}M_{j}\rightarrow A$. $M_{j}$ is unlimited except that if a
mechanism is direct, \emph{i.e.}, $M_{j}=T_{j}$.

An SCR $F$ satisfies \emph{no-veto} if, whenever $a\succeq_{j}^{t}b$
for all $b\in A$ and for every agent $j$ but perhaps one $k$, then
$a\in F(E)$. An SCR $F$ is \emph{monotonic} if for every pair of
environments $E$ and $E'$, and for every $a\in F(E)$, whenever
$a\succeq_{j}^{t}b$ implies that $a\succeq_{j}^{t'}b$, there holds
$a\in F(E')$. We assume that there is \emph{complete information}
among the agents, \emph{i.e.}, the true state $t$ is common
knowledge among them. Given a mechanism $\Gamma=((M_{j})_{j\in
N},g)$ played in state $t$, a \emph{Nash equilibrium} of $\Gamma$ in
state $t$ is a strategy profile $m^{*}$ such that: $\forall j\in N,
g(m^{*}(t))\succeq_{j}^{t}g(m_{j},m_{-j}^{*}(t)), \forall m_{j}\in
M_{j}$. Let $\mathcal {N}(\Gamma,t)$ denote the set of Nash
equilibria of the game induced by $\Gamma$ in state $t$, and
$g(\mathcal {N}(\Gamma,t))$ denote the corresponding set of Nash
equilibrium outcomes. An SCR $F$ is \emph{Nash implementable} if
there exists a mechanism $\Gamma=((M_{j})_{j\in N},g)$ such that for
every $t\in \mathcal {T}$, $g(\mathcal {N}(\Gamma,t))=F(t)$.

Maskin \cite{Maskin1999} provided an almost complete
characterization of SCRs that were Nash implementable. The main
results of Ref. \cite{Maskin1999} are two theorems: 1)
(\emph{Necessity}) If an SCR is Nash implementable, then it is
monotonic. 2) (\emph{Sufficiency}) Let $n\geq3$, if an SCR is
monotonic and satisfies no-veto, then it is Nash implementable. In
order to facilitate the following investigation, we briefly recall
the Maskin mechanism given by Serrano \cite{Serrano2004} as follows:

Consider a mechanism $\Gamma=((M_{j})_{j\in N},g)$, where agent
$j$'s message set is $M_{j}=A\times \mathcal {T} \times
\mathbb{Z}_{+}$, $\mathbb{Z}_{+}$ is the set of non-negative
integers. A typical message sent by agent $j$ is described as
$m_{j}=(a_{j},t_{j},z_{j})$. The outcome function $g$ is defined in
the following three rules: (1) If for every agent $j\in N$,
$m_{j}=(a,t,0)$ and $a\in F(t)$, then $g(m)=a$. (2) If $(n-1)$
agents $j\neq k$ send $m_{j}=(a,t,0)$ and $a\in F(t)$, but agent $k$
sends $m_{k}=(a_{k},t_{k},z_{k})\neq(a,t,0)$, then $g(m)=a$ if
$a_{k}\succ_{k}^{t}a$, and $g(m)=a_{k}$ otherwise. (3) In all other
cases, $g(m)=a'$, where $a'$ is the outcome chosen by the agent with
the lowest index among those who announce the highest integer.

\section{Nash implementation on Internet}
This section is the main part of this paper. In the beginning, we
will show an example of SCR which satisfies monotonicity and
no-veto. It is Nash implementable although all agents dislike it.
Then, we will propose a self-enforcing agreement using complex
numbers, by which the agents may break through the Maskin's
sufficiency theorem and make the SCR not Nash implementable.

\subsection{A Pareto-inefficient SCR that satisfies monotonicity and no-veto}
Let $N=\{Apple, Lily, Cindy\}$, $\mathcal {T}=\{t^{1},t^{2}\}$,
$A=\{a^{1},a^{2},a^{3},a^{4}\}$. In each state $t\in\mathcal {T}$,
the preference relations $(\succeq^{t}_{j})_{j\in N}$ over the
outcome set $A$ and the corresponding SCR $F$ are given in Table 1.
The SCR $F$ is \emph{Pareto-inefficient} from the agents'
perspectives because in state $t^{2}$, all agents unanimously prefer
a Pareto-optimal outcome $a^{1}$: for each agent $j\in N$,
$a^{1}\succ^{t^{2}}_{j}a^{2}\in F(t^{2})$.

\emph{Table 1: An SCR satisfying monotonicity and no-veto is Pareto-inefficient
from the agents' perspectives.}\\
\begin{tabular}{cccccc}
 \multicolumn{3}{c}{State $t^{1}$}&\multicolumn{3}{c}{State $t^{2}$}\\
 $Apple$&$Lily$ &$Cindy$ &\quad$Apple$&$Lily$ &$Cindy$\\ \hline
 $a^{3}$&$a^{2}$&$a^{1}$ &$a^{4}$&$a^{3}$&$a^{1}$ \\
 $a^{1}$&$a^{1}$&$a^{3}$ &$a^{1}$&$a^{1}$&$a^{2}$ \\
 $a^{2}$&$a^{4}$&$a^{2}$ &$a^{2}$&$a^{2}$&$a^{3}$ \\
 $a^{4}$&$a^{3}$&$a^{4}$ &$a^{3}$&$a^{4}$&$a^{4}$ \\\hline
 \multicolumn{3}{c}{$F(t^{1})=\{a^{1}\}$}&\multicolumn{3}{c}{$F(t^{2})=\{a^{2}\}$}\\\hline
\end{tabular}

Suppose the true state is $t^{2}$. At first sight, $(a^{1},t^{1},0)$
might be a unanimous $m_{j}$ for each agent $j$, because by doing so
$a^{1}$ would be generated by rule 1 of the Maskin mechanism.
However, $Apple$ has an incentive to unilaterally deviate from
$(a^{1},t^{1},0)$ to $(a^{4},*,*)$ in order to trigger rule 2 (where
$*$ stands for any legal value), since
$a^{1}\succ^{t^{1}}_{Apple}a^{4}$,
$a^{4}\succ^{t^{2}}_{Apple}a^{1}$; $Lily$ also has an incentive to
unilaterally deviate from $(a^{1},t^{1},0)$ to $(a^{3},*,*)$, since
$a^{1}\succ^{t^{1}}_{Lily}a^{3}$, $a^{3}\succ^{t^{2}}_{Lily}a^{1}$.

Note that either $Apple$ or $Lily$ can certainly obtain her expected
outcome only if just one of them deviates from $(a^{1},t^{1},0)$ (If
this case happened, rule 2 would be triggered). But this condition
is unreasonable, because all agents are rational, nobody is willing
to give up and let the others benefit. Therefore, both $Apple$ and
$Lily$ will deviate from $(a^{1},t^{1},0)$. As a result, rule 3 will
be triggered. Since $Apple$ and $Lily$ both have a chance to win the
integer game, the winner is uncertain and the final outcome is also
uncertain between $a^{3}$ and $a^{4}$.

To sum up, although every agent prefers $a^{1}$ to $a^{2}$ in state
$t^{2}$, $a^{1}$ cannot be yielded in Nash equilibrium. Indeed, the
Maskin mechanism makes the Pareto-inefficient outcome $a^{2}$ be
Nash implementable in state $t^{2}$.

Can the agents find a way to break through the Maskin's sufficiency
theorem and let the Pareto-efficient outcome $a^{1}$ be Nash
implementable in state $t^{2}$? Interestingly, we will show that the
answer may be ``yes'' if agents report messages to the designer
through channels (\emph{e.g.}, Internet). In what follows, first we
will define some matrices with complex numbers, then we will propose
a self-enforcing agreement to help agents break through the Maskin's
sufficiency theorem.

\subsection{Definitions}
\textbf{Definition 1}: Let $\hat{I}, \hat{\sigma}$ be two $2\times
2$ matrices, and $\overrightarrow{C}, \overrightarrow{D}$ be two
basis vectors:
\begin{equation} \quad \hat{I}\equiv\begin{bmatrix}
  1 & 0\\
  0 & 1
\end{bmatrix},\quad \hat{\sigma}\equiv\begin{bmatrix}
  0 & 1\\
  1 & 0
\end{bmatrix}, \overrightarrow{C}\equiv\begin{bmatrix}
  1\\
  0
\end{bmatrix},\quad \overrightarrow{D}\equiv\begin{bmatrix}
  0\\
  1
\end{bmatrix}.
\end{equation}
Hence, $\hat{I}\overrightarrow{C}=\overrightarrow{C}$,
$\hat{I}\overrightarrow{D}=\overrightarrow{D}$;
$\hat{\sigma}\overrightarrow{C}=\overrightarrow{D}$,
$\hat{\sigma}\overrightarrow{D}=\overrightarrow{C}$.

\textbf{Definition 2}: For $n\geq 3$ agents, suppose each agent
$j\in N$ possesses a basis vector. $\overrightarrow{\psi}_{0}$ is
defined as the tensor product of $n$ basis vectors
$\overrightarrow{C}$:
\begin{equation}
\overrightarrow{\psi}_{0}\equiv\overrightarrow{C}^{\otimes
n}\equiv\underbrace{\overrightarrow{C}\otimes\cdots \otimes
\overrightarrow{C}}\limits_{n}\equiv\begin{bmatrix}
  1\\
  0\\
  \cdots\\
  0
\end{bmatrix}_{2^{n}\times 1}
\end{equation}
$\overrightarrow{C}^{\otimes n}$ contains $n$ basis vectors
$\overrightarrow{C}$ and $2^{n}$ elements.
$\overrightarrow{C}^{\otimes n}$ is also denoted as
$\overrightarrow{C\cdots CC}^{n}$. Similarly,
\begin{equation}
\overrightarrow{C\cdots
CD}^{n}\equiv\underbrace{\overrightarrow{C}\otimes\cdots \otimes
\overrightarrow{C}}\limits_{n-1}\otimes
\overrightarrow{D}=\begin{bmatrix}
  0\\
  1\\
  \cdots\\
  0
\end{bmatrix}_{2^{n}\times 1}
\end{equation}
Obviously, there are $2^{n}$ possible vectors:
$\overrightarrow{C\cdots CC}^{n},$ $ \cdots,\overrightarrow{D\cdots
DD}^{n}$.

\textbf{Definition 3}: $\hat{J}\equiv
\frac{1}{\sqrt{2}}(\hat{I}^{\otimes n}+i\hat{\sigma}^{\otimes  n})$,
\emph{i.e.},
\begin{equation}
\hat{J}\equiv\frac{1}{\sqrt{2}}\begin{bmatrix}
  1 &  &  &  &  &  & i\\
   & \cdots  & &  & & \cdots  & \\
   &  &  & 1 & i &  & \\
   &  &  & i & 1 &  & \\
   & \cdots  & &  &  & \cdots & \\
  i &  &  &  &  &  & 1
\end{bmatrix}_{2^{n}\times2^{n}},
\hat{J}^{+}\equiv\frac{1}{\sqrt{2}}\begin{bmatrix}
  1 &  &  &  &  &  & -i\\
   & \cdots  & &  & & \cdots  & \\
   &  &  & 1 & -i &  & \\
   &  &  & -i & 1 &  & \\
   & \cdots  & &  &  & \cdots & \\
  -i &  &  &  &  &  & 1
\end{bmatrix}_{2^{n}\times2^{n}}
\end{equation}
where the symbol $i$ denotes an imaginary number, and $\hat{J}^{+}$
is the conjugate transpose of $\hat{J}$.

\textbf{Definition 4}:
\begin{equation}
\overrightarrow{\psi}_{1}\equiv
\hat{J}\overrightarrow{\psi}_{0}=\frac{1}{\sqrt{2}}\begin{bmatrix}
  1\\
  0\\
  \cdots\\
  0\\
  i
\end{bmatrix}_{2^{n}\times 1}
\end{equation}
\textbf{Definition 5}: For $\theta\in[0, \pi]$, $\phi\in [0,
\pi/2]$,
\begin{equation}
  \hat{\omega}(\theta, \phi)\equiv\begin{bmatrix}
  e^{i\phi}\cos(\theta/2) & i\sin(\theta/2)\\
  i\sin(\theta/2) & e^{-i\phi}\cos(\theta/2)
\end{bmatrix}.
\end{equation}
$\hat{\Omega}\equiv\{\hat{\omega}(\theta,\phi):\theta\in[0,\pi],\phi\in[0,\pi/2]\}$.
Hence, $\hat{I}=\hat{\omega}(0, 0)$.

 \textbf{Definition 6}: For $j=1, \cdots, n$, $\theta_{j}\in[0, \pi]$, $\phi_{j}\in [0,
\pi/2]$, let  $\hat{\omega}_{j}=\hat{\omega}(\theta_{j}, \phi_{j})$,
\begin{equation}
\overrightarrow{\psi}_{2}\equiv[\hat{\omega}_{1}\otimes\cdots\otimes\hat{\omega}_{n}]\overrightarrow{\psi}_{1}.
\end{equation}
The dimension of
$\hat{\omega}_{1}\otimes\cdots\otimes\hat{\omega}_{n}$ is
$2^{n}\times 2^{n}$. Since only two elements in
$\overrightarrow{\psi}_{1}$ are non-zero, it is not necessary to
calculate the whole $2^{n}\times2^{n}$ matrix to yield
$\overrightarrow{\psi}_{2}$. Indeed, we only need to calculate the
leftmost and rightmost column of
$\hat{\omega}_{1}\otimes\cdots\otimes\hat{\omega}_{n}$ to derive
$\overrightarrow{\psi}_{2}$.

\textbf{Definition 7}: $\overrightarrow{\psi}_{3}\equiv
\hat{J}^{+}\overrightarrow{\psi}_{2}$.

Suppose $\overrightarrow{\psi}_{3}=[\eta_{1}, \cdots,
\eta_{2^{n}}]^{T}$, let $\Delta=[|\eta_{1}|^{2}, \cdots,
|\eta_{2^{n}}|^{2}]$. It can be easily checked that $\hat{J}$,
$\hat{\omega}_{j}$ ($j=1,\cdots, n$) and $\hat{J}^{+}$ are all
unitary matrices. Hence, $|\overrightarrow{\psi}_{3}|^{2}=1$. Thus,
$\Delta$ can be viewed as a probability distribution, each element
of which represents the probability that we randomly choose a vector
from the set of all $2^{n}$ possible vectors
$\{\overrightarrow{C\cdots CC}^{n},$ $
\cdots,\overrightarrow{D\cdots DD}^{n}\}$.

\textbf{Definition 8}: Condition $\lambda$ contains
five parts. The first three parts are defined as follows:\\
$\lambda_{1}$: Given an SCR $F$, there exist two states $\hat{t}$,
$\bar{t}\in \mathcal {T}$, $\hat{t}\neq \bar{t}$ such that
$\hat{a}\succeq^{\bar{t}}_{j}\bar{a}$ (for each $j\in N$,
$\hat{a}\in F(\hat{t})$, $\bar{a}\in F(\bar{t})$) with strict
relation for some agent; and the number of agents that encounter a
preference change around $\hat{a}$ in going from state $\hat{t}$ to
$\bar{t}$ is at least two. Denote by $\hat{N}$ the set of these
agents and $l=|\hat{N}|$ the number of these agents.\\
$\lambda_{2}$: Consider the state $\bar{t}$ specified in condition
$\lambda_{1}$, if there exists another $\hat{t}'\in \mathcal {T}$,
$\hat{t}'\neq\hat{t}$ that satisfies $\lambda_{1}$, then
$\hat{a}\succeq^{\bar{t}}_{j}\hat{a}'$ (for each $j\in N$,
$\hat{a}\in F(\hat{t})$, $\hat{a}'\in
F(\hat{t}')$) with strict relation for some agent.\\
$\lambda_{3}$: Consider the outcome $\hat{a}$ specified in condition
$\lambda_{1}$, for any state $t\in\mathcal{T}$, $\hat{a}$ is top
ranked for each agent $j\in N\backslash\hat{N}$.

\subsection{An agreement using complex numbers}
As specified before, in this paper we will discuss how the
traditional results on Nash implementation will be changed when
agents report messages to the designer through channels. We assume
that:

1) Each agent has a unique channel connected with the designer. The
agents report messages to the designer through these channels.

2) After the designer claims the outcome function $g$, if all agents
anticipate the SCR $F$ appeared in rule 1 of $g$ satisfies condition
$\lambda$ (given in Definition 8 and 11), agents can negotiate and
construct an agreement \emph{ComplexMessage} as shown in Fig. 1. The
algorithm \emph{MessageComputing} is given in Definition 9. In the
initial configuration, the computer controls all channels.

3) Each agent can freely decide whether to leave his channel to the
computer and let the computer send a message to the designer, or to
take back his channel and send his message to the designer by
himself. If any agent deviates from the initial configuration and
chooses the latter option, then this deviation is \emph{observable}
to the rest agents and all of them will choose the latter option
too, \emph{i.e.}, all agents will take back their channels and send
their messages to the designer by themselves. If all agents leave
their channels to the computer, the algorithm
\emph{MessageComputing} works, \emph{i.e.}, calculates $m_{1},
\cdots, m_{n}$ and sends them to the designer through channels.

\begin{figure}[!t]
\centering
\includegraphics[height=2.8in,clip,keepaspectratio]{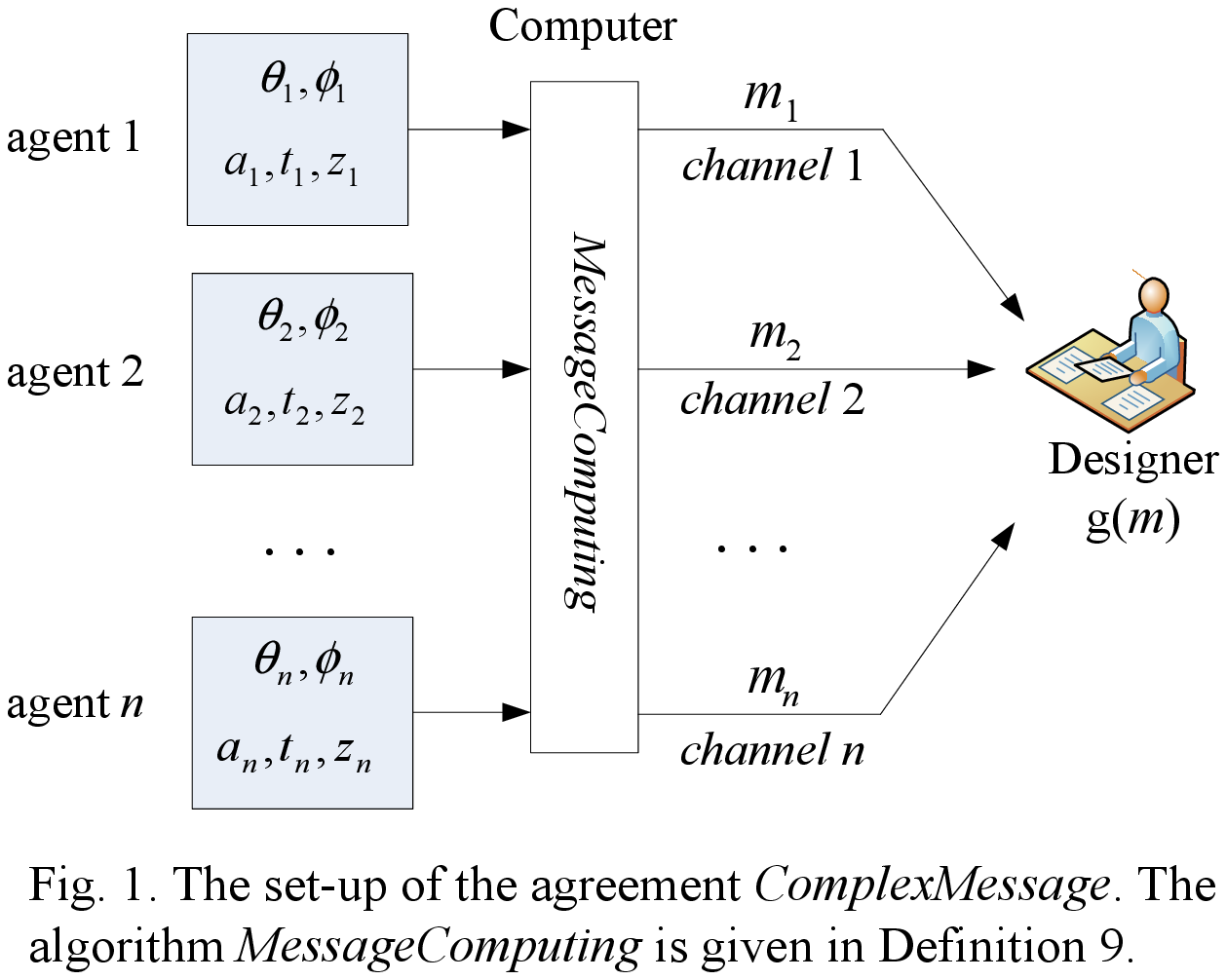}
\end{figure}

\textbf{Definition 9}: The algorithm \emph{MessageComputing} is defined as follows:\\
\textbf{Input}: $(\theta_{j}, \phi_{j}, a_{j},t_{j},z_{j})\in
[0,\pi/2]\times [0,\pi]\times A\times \mathcal {T} \times
\mathbb{Z}_{+}$,
$j=1,\cdots,n$.\\
\textbf{Output}:  $m_{j}\in A\times \mathcal {T} \times
\mathbb{Z}_{+}$, $j=1,\cdots,n$.\\
Step 1: Reading $(\theta_{j},
\phi_{j})$ from each agent $j\in N$.\\
Step 2: Computing the leftmost and rightmost columns of
$\hat{\omega}_{1}\otimes\cdots\otimes\hat{\omega}_{n}$.\\
Step 3: Computing
$\overrightarrow{\psi}_{2}=[\hat{\omega}_{1}\otimes\cdots\otimes\hat{\omega}_{n}]\overrightarrow{\psi}_{1}$,
$\overrightarrow{\psi}_{3}=\hat{J}^{+}\overrightarrow{\psi}_{2}$,
and the probability distribution
$\Delta$.\\
Step 4: Randomly choosing a vector from the set of all $2^{n}$
possible vectors $\{\overrightarrow{C\cdots CC}^{n},$ $
\cdots,\overrightarrow{D\cdots DD}^{n}\}$ according to the
probability distribution $\Delta$.\\
Step 5: For each agent $j\in N$, let $m_{j}=(\hat{a}, \hat{t}, 0)$
(or $m_{j}=(a_{j},t_{j},z_{j})$) if the $j$-th basis vector of the
chosen vector is $\overrightarrow{C}$ (or $\overrightarrow{D}$),
where $\hat{a}, \hat{t}$
are specified in condition $\lambda_{1}$.\\
Step 6: Sending $m=(m_{1}, \cdots, m_{n})$ to the designer through
channels $1, \cdots, n$.

\textbf{Definition 10}: Suppose $\lambda_{1}$ and $\lambda_{2}$ are
satisfied and $m=(m_{1}, \cdots, m_{n})$ is computed by
\emph{MessageComputing}. Suppose the true state is $\bar{t}$
specified in condition $\lambda_{1}$. $\$_{C\cdots CC}$,
$\$_{C\cdots CD}$, $\$_{D\cdots DC}$ and $\$_{D\cdots DD}$ are
defined as the payoffs to the $n$-th agent in state $\bar{t}$ when
the chosen vector in Step 4 of \emph{MessageComputing} is
$\overrightarrow{C\cdots CC}^{n}$, $\overrightarrow{C\cdots
CD}^{n}$, $\overrightarrow{D\cdots DC}^{n}$ or
$\overrightarrow{D\cdots DD}^{n}$ respectively.

\textbf{Definition 11}: Suppose conditions $\lambda_{1}$,
$\lambda_{2}$ and $\lambda_{3}$ are satisfied and the true state is
$\bar{t}$, consider each message $m_{j}=(a_{j}, t_{j}, z_{j})$,
where $a_{j}$ is top-ranked for each agent $j$. The rest two parts
of condition
$\lambda$ are defined as:\\
$\lambda_{4}$: For each agent $j\in N$, let him be the $n$-th
agent, $\$_{C\cdots CC}>\$_{D\cdots DD}$.\\
$\lambda_{5}$: For each agent $j\in \hat{N}$, let him be the $n$-th
agent, $\$_{C\cdots CC}>\$_{C\cdots
CD}\cos^{2}(\pi/l)+\$_{D\cdots DC}\sin^{2}(\pi/l)$.\\

It can be seen from Fig. 1 that after the agreement
\emph{ComplexMessage} is
constructed, each agent $j\in N$ independently faces two options:\\
$\bullet$ $S(j,0)$: leaving his channel to the computer, and
submitting
$(\theta_{j}, \phi_{j}, a_{j}, t_{j}, z_{j})$ to the algorithm \emph{MessageComputing}.\\
$\bullet$ $S(j,1)$: taking back his channel, and submitting $(a_{j},
t_{j}, z_{j})$ to the designer by himself.

To sum up, if agents report messages to the designer through
channels and all agents anticipate the SCR $F$ appeared in rule 1 of
$g$ satisfies condition $\lambda$, they can construct the agreement
\emph{ComplexMessage} after the designer claims the outcome function
$g$. The timing steps of the Maskin mechanism are updated as follows:\\
Time 1: The designer claims the outcome function $g$ to all agents;\\
Time 2: The agents construct the agreement \emph{ComplexMessage};\\
Time 3: Each agent $j\in N$ chooses an option between $S(j,0)$ and
$S(j,1)$, and $m_{1}, \cdots, m_{n}$ are sent through channels.\\
Time 4: The designer receives $m=(m_{1}, \cdots, m_{n})$ from $n$ channels; \\
Time 5: The designer announces the outcome $g(m)$.

\subsection{Main result}
\textbf{Proposition 1}: For $n\geq 3$, suppose agents send messages
to the designer through channels (\emph{e.g.}, Internet). Consider
an SCR $F$ that satisfies monotonicity and no-veto. If condition
$\lambda$ is satisfied, then in state $\bar{t}$ the agents can
construct a self-enforcing agreement \emph{ComplexMessage} to make
the Pareto-inefficient outcome $F(\bar{t})$ not be implemented in
Nash equilibrium.

\textbf{Proof}: Since $\lambda_{1}$ and $\lambda_{2}$ are satisfied,
then there exist two states $\hat{t}$, $\bar{t}\in \mathcal {T}$,
$\hat{t}\neq \bar{t}$ such that
$\hat{a}\succeq^{\bar{t}}_{j}\bar{a}$ (for each $j\in N$,
$\hat{a}\in F(\hat{t})$, $\bar{a}\in F(\bar{t})$) with strict
relation for some agent, and $\hat{N}$ contains the agents that
encounter a preference change around $\hat{a}$ in going from state
$\hat{t}$ to $\bar{t}$ ($l=|\hat{N}|\geq 2$). Suppose the true state
is $\bar{t}$, now let us check what will happen after the agents
construct the agreement \emph{ComplexMessage} in Time 2.

From the viewpoints of agents, after constructing
\emph{ComplexMessage},
there are two possible cases in Time 3:\\
1) Suppose every agent $j\in N$ chooses $S(j,0)$, then the algorithm
\emph{MessageComputing} works. Consider the following strategy
profile: each agent $j\in N\backslash\hat{N}$ submits $(\theta_{j},
\phi_{j})=(0, 0)$, $(a_{j}, t_{j}, z_{j})=(\hat{a}, \hat{t}, 0)$;
each agent $j\in\hat{N}$ submits $(\theta_{j}, \phi_{j})=(0,\pi/l)$.
According to Lemma 1 (see Appendix), this strategy profile is a Nash
equilibrium of $\Gamma$ in state $\bar{t}$. In Step 4 of
\emph{MessageComputing}, the chosen vector is
$\overrightarrow{C\cdots CC}$. In Step 5 of \emph{MessageComputing},
$m_{j}=(\hat{a}, \hat{t}, 0)$ for each $j\in N$. Therefore, in Time
5, $g(m)=\hat{a}\notin F(\bar{t})$. Each agent $j$'s payoff is
$\$_{C\cdots CC}$.\\
2) Suppose some agent $j\in N$ chooses $S(j,1)$, \emph{i.e.}, takes
back his channel and reports $m_{j}$ to the designer by himself.
Then according to assumption 3 (see Section 3.3), all of the rest
agents will observe this deviation, thereby take back their channels
and submit messages to the designer by themselves. According the
Maskin mechanism, in Time 5, the final outcome will be either
$F(\bar{t})$ by rule 1, or uncertain by rule 3 (\emph{i.e.}, each
agent $j$'s payoff is $\$_{D\cdots DD}$).

According to conditions $\lambda_{1}$ and $\lambda_{4}$, it is not
profitable for any agent $j$ to choose $S(j,1)$, \emph{i.e.},
unilaterally take back his channel and send a message to the
designer by himself. As Telser pointed out in page 28, Line 2
\cite{Telser1980}: ``\emph{A party to a self-enforcing agreement
calculates whether his gain from violating the agreement is greater
or less than the loss of future net benefits that he would incur as
a result of detection of his violation and the consequent
termination of the agreement by the other party... Hence both
parties continue to adhere to an agreement if and only if each gains
more from adherence to, than from violations of, its terms.}''
Therefore, it can be seen that \emph{ComplexMessage} is a
self-enforcing agreement among the agents. Put differently, although
the agents collaborate to construct \emph{ComplexMessage} in Time 2,
they do not require a third-party to enforce it after then and the
game is still non-cooperative.

To sum up, in state $\bar{t}$, the agents can construct a
self-enforcing agreement \emph{ComplexMessage} to make the
Pareto-inefficient outcome $F(\bar{t})$ not be implemented in Nash
equilibrium.\quad
\quad\quad\quad\quad\quad\quad\quad\quad\quad\quad\quad\quad
\quad\quad\quad\quad\quad\quad\quad\quad\quad\quad\quad$\square$

Let us reconsider Table 1. Let $\hat{t}=t^{1}$, $\hat{a}=a^{1}$,
$\bar{t}=t^{2}$, $\bar{a}=a^{2}$. Suppose the true state is $t^{2}$.
Since both $Apple$ and $Lily$ encounter a preference change around
$a^{1}$ in going from state $t^{1}$ to $t^{2}$, condition
$\lambda_{1}$ is satisfied. Obviously, $\lambda_{2}$ and
$\lambda_{3}$ are also satisfied. Consider the strategy profile as
follows:
\begin{align*}
&(\theta_{Apple}, \phi_{Apple})=(0, \pi/2), \quad&&(a_{Apple},
t_{Apple},
z_{Apple})=(a^{4}, *, *);\\
&(\theta_{Lily}, \phi_{Lily})=(0, \pi/2), \quad&&(a_{Lily},
t_{Lily},
z_{Lily})=(a^{3}, *, *);\\
&(\theta_{Cindy}, \phi_{Cindy})=(0, 0), \quad&&(a_{Cindy},
t_{Cindy}, z_{Cindy})=(a^{1}, t^{1}, 0).
\end{align*}
Let $Cindy$ be the first agent, and for any agent $j\in\{Apple,
Lily\}$, let her be the last agent. Consider the payoff to the third
agent, suppose $\$_{CCC}=3$ (the corresponding outcome is $a^{1}$),
$\$_{CCD}=5$ (the corresponding outcome is $a^{4}$ if $j=Apple$, and
$a^{3}$ if $j=Lily$), $\$_{DDC}=0$ (the corresponding outcome is
$a^{3}$ if $j=Apple$, and $a^{4}$ if $j=Lily$), $\$_{DDD}=1$ (the
corresponding outcome is uncertain between $a^{3}$ and $a^{4}$). Let
$Cindy$ be the last agent and consider her payoff, suppose
$\$_{CCC}=3$ and $\$_{DDD}=1$. Hence, $\lambda_{4}$ and
$\lambda_{5}$ are satisfied. According to Proposition 1, in state
$t^{2}$, the outcome implemented in Nash equilibrium is $a^{1}$, and
$F(t^{2})$ is not Nash implementable although the SCR $F$ satisfies
monotonicity and no-veto.

\textbf{Remark 1:} Some reader may argue that the agreement
\emph{ComplexMessage} is a wrapper to the Maskin mechanism that
changes the game substantially and henceforth it has no implication
on the original Maskin mechanism. However, this viewpoint is not
true. Actually, \emph{ComplexMessage} is \emph{unobservable} to the
designer because it is hidden behind channels and the designer
cannot prevent the agents from constructing such agreement. From the
designer's perspective, no matter whether the agents construct the
agreement \emph{ComplexMessage} on the other side of channels or
not, the Maskin mechanism remain unchanged and the designer acts in
the same way: \emph{i.e.}, claims the outcome function $g$, receives
messages $m=(m_{1},\cdots,m_{n})$ from channels and announces the
final outcome $g(m)$.

\textbf{Remark 2:} Although the time and space complexity of
\emph{MessageComputing} are exponential with the number of agents,
\emph{i.e.}, $O(2^{n})$, it works well when the number of agents is
not very large. For example, the runtime of \emph{MessageComputing}
is about 0.5s for 15 agents, and about 12s for 20 agents (MATLAB
7.1, CPU: Intel (R) 2GHz, RAM: 3GB).

\textbf{Remark 3:} The problem of Nash implementation requires
complete information among all agents. In the last paragraph of Page
392 \cite{Serrano2004}, Serrano wrote: ``\emph{We assume that there
is complete information among the agents... This assumption is
especially justified when the implementation problem concerns a
small number of agents that hold good information about one
another}''. Hence, the fact that \emph{MessageComputing} is suitable
for small-scale cases (\emph{e.g.}, less than 20 agents) is
acceptable for Nash implementation.

\section{Conclusion}
In this paper, we propose a self-enforcing agreement to help agents
avoid a Pareto-inefficient social choice rule if agents report
messages to the designer through channels and condition $\lambda$ is
satisfied. Put differently, traditional sufficient conditions for
Nash implementation may fail on Internet. With the rapid development
of network economics, it will be more and more common that agents
communicate with the designer through Internet. In the future, there
are many works to do to study the self-enforcing agreement further.

\section*{Acknowledgments}

The author is very grateful to Ms. Fang Chen, Hanyue Wu
(\emph{Apple}), Hanxing Wu (\emph{Lily}) and Hanchen Wu
(\emph{Cindy}) for their great support.

%--------------------------------------------------------

%--------------------------------------------------------
\newpage
\section*{Appendix}
\textbf{Lemma 1}: Suppose condition $\lambda$ is satisfied and the
algorithm \emph{MessageComputing} works. Consider the following strategy
profile:\\
1) Each agent $j\in N\backslash\hat{N}$ submits $(\theta_{j},
\phi_{j})=(0, 0)$, $(a_{j}, t_{j}, z_{j})=(\hat{a}, \hat{t}, 0)$; \\
2) Each agent $j\in\hat{N}$ submits $(\theta_{j},
\phi_{j})=(0,\pi/l)$;\\
then this strategy profile is a Nash equilibrium of $\Gamma$ in
state $\bar{t}$, where $\bar{t}$ is specified in condition
$\lambda_{1}$.

\textbf{Proof}: The proof consists of two parts.

\emph{Part 1}. Let the last $l$ agents be $\hat{N}$. Consider the
following strategy profile: each agent $j=1, \cdots, (n-l)$ submits
$(\theta_{j}, \phi_{j})=(0, 0)$, $(a_{j}, t_{j}, z_{j})=(\hat{a},
\hat{t}, 0)$; each agent $j=(n-l+1), \cdots, (n-1)$ submits
$(\theta_{j}, \phi_{j})=(0,\pi/l)$, then we will prove the optimal
value of $(\theta_{n},\phi_{n})$ for the $n$-th agent is
$(0,\pi/l)$.

Since condition $\lambda_{1}$ is satisfied, then $l\geq 2$. Let
\begin{equation*}
  \hat{C}_{l}\equiv\hat{\omega}(0,\pi/l)=\begin{bmatrix}
  e^{i\frac{\pi}{l}} & 0 \\
  0 & e^{-i\frac{\pi}{l}}
\end{bmatrix}_{2\times 2},\quad\mbox{ thus, }
  \hat{C}_{l}\otimes\hat{C}_{l}=\begin{bmatrix}
  e^{i\frac{2\pi}{l}} &   &   & \\
  & 1 &   & \\
  &   &1  & \\
  & &  & e^{-i\frac{2\pi}{l}}
\end{bmatrix}_{2^{2}\times 2^{2}},
\end{equation*}
\begin{equation*}
\underbrace{\hat{C}_{l}\otimes\cdots \otimes
\hat{C}_{l}}\limits_{l-1}=\begin{bmatrix}
  e^{i\frac{(l-1)}{l}\pi} &   &   & \\
  & * &   & \\
  &   &\cdots  & \\
  & &  & e^{-i\frac{(l-1)}{l}\pi}
\end{bmatrix}_{2^{l-1}\times 2^{l-1}}.
\end{equation*}
Here we only explicitly show the up-left and bottom-right entries
because only these two entries are useful in the following
discussions. The other entries in diagonal are simply represented as
symbol $*$. Note that
\begin{equation*}
\underbrace{\hat{I}\otimes\cdots \otimes
\hat{I}}\limits_{n-l}=\begin{bmatrix}
  1 &   &   & \\
  & 1 &   & \\
  &   &\cdots  & \\
  & &  & 1
\end{bmatrix}_{2^{n-l}\times 2^{n-l}},
\end{equation*}
thus,
\begin{equation*}
\underbrace{\hat{I}\otimes\cdots \otimes
\hat{I}}\limits_{n-l}\otimes\underbrace{\hat{C}_{l}\otimes\cdots
\otimes \hat{C}_{l}}\limits_{l-1}=\begin{bmatrix}
  e^{i\frac{(l-1)}{l}\pi} &   &   & \\
  & * &   & \\
  &   &\cdots  & \\
  & &  & e^{-i\frac{(l-1)}{l}\pi}
\end{bmatrix}_{2^{n-1}\times 2^{n-1}}.
\end{equation*}
Suppose the $n$-th agent chooses arbitrary parameters $(\theta,
\phi)$ in his strategy $(\theta, \phi, a_{n}, t_{n}, z_{n})$, let
\begin{equation*}
  \hat{\omega}_{n}(\theta, \phi)=\begin{bmatrix}
  e^{i\phi}\cos(\theta/2) & i\sin(\theta/2)\\
  i\sin(\theta/2) & e^{-i\phi}\cos(\theta/2)
\end{bmatrix},
\end{equation*}
then,
\begin{align*}
\underbrace{\hat{I}\otimes\cdots \otimes
\hat{I}}\limits_{n-l}&\otimes\underbrace{\hat{C}_{l}\otimes\cdots
\otimes \hat{C}_{l}}\limits_{l-1}\otimes\hat{\omega}_{n}(\theta,
\phi)\\
&=\begin{bmatrix}
  e^{i[\frac{(l-1)\pi}{l}+\phi]}\cos(\theta/2) &*   &   &   &   &    & \\
  ie^{i\frac{(l-1)\pi}{l}}\sin(\theta/2) &*   &   &   &   &    & \\
   &  & *  & *  &   &    & \\
   &  & *  & *  &   &    & \\
   &  &    &    &\cdots&    & \\
   &  &    &    & & * & ie^{-i\frac{(l-1)\pi}{l}}\sin(\theta/2) \\
   &  &    &    & & * & e^{-i[\frac{(l-1)\pi}{l}+\phi]}\cos(\theta/2)
\end{bmatrix}_{2^{n}\times 2^{n}}.
\end{align*}
Recall that
\begin{equation*}
\overrightarrow{\psi}_{1}=\frac{1}{\sqrt{2}}\begin{bmatrix}
  1\\
  0\\
  \cdots\\
  0\\
  i
\end{bmatrix}_{2^{n}\times 1},
\end{equation*}
thus,
\begin{align*}
\overrightarrow{\psi}_{2}=[\underbrace{\hat{I}\otimes\cdots \otimes
\hat{I}}\limits_{n-l}\otimes\underbrace{\hat{C}_{l}\otimes\cdots
\otimes \hat{C}_{l}}\limits_{l-1}\otimes\hat{\omega}_{n}(\theta,
\phi)]\overrightarrow{\psi}_{1}=\frac{1}{\sqrt{2}}\begin{bmatrix}
  e^{i[\frac{(l-1)\pi}{l}+\phi]}\cos(\theta/2)\\
  ie^{i\frac{(l-1)\pi}{l}}\sin(\theta/2)\\
  0\\
  \cdots\\
  0\\
  -e^{-i\frac{(l-1)\pi}{l}}\sin(\theta/2)\\
  ie^{-i[\frac{(l-1)\pi}{l}+\phi]}\cos(\theta/2)
\end{bmatrix}_{2^{n}\times 1},
\end{align*}
\begin{align*}
\overrightarrow{\psi}_{3}=\hat{J}^{+}\overrightarrow{\psi}_{2}=\begin{bmatrix}
  \cos(\theta/2)\cos(\frac{l-1}{l}\pi+\phi)\\
  i\sin(\theta/2)\cos\frac{l-1}{l}\pi\\
  0\\
  \cdots\\
  0\\
  i\sin(\theta/2)\sin\frac{l-1}{l}\pi \\
  \cos(\theta/2)\sin(\frac{l-1}{l}\pi+\phi)
\end{bmatrix}_{2^{n}\times 1}.
\end{align*}
The probability distribution $\Delta$ is computed from
$\overrightarrow{\psi}_{3}$:
\begin{align*}
&P_{C\cdots CC}=\cos^{2}(\theta/2)\cos^{2}(\phi-\frac{\pi}{l})\\
&P_{C\cdots CD}=\sin^{2}(\theta/2)\cos^{2}\frac{\pi}{l}\\
&P_{D\cdots DC}=\sin^{2}(\theta/2)\sin^{2}\frac{\pi}{l}\\
&P_{D\cdots DD}=\cos^{2}(\theta/2)\sin^{2}(\phi-\frac{\pi}{l})
\end{align*}
Obviously,
\begin{equation*}
P_{C\cdots CC}+P_{C\cdots CD}+P_{D\cdots DC}+P_{D\cdots DD}=1.
\end{equation*}
Consider the payoff to the $n$-th agent,
\begin{equation*}
\$_{n}=\$_{C\cdots CC}P_{C\cdots CC}+\$_{C\cdots CD}P_{C\cdots CD}
  +\$_{D\cdots DC}P_{D\cdots DC}+\$_{D\cdots DD}P_{D\cdots DD}.
\end{equation*}
Since $\lambda_{4}$ is satisfied, \emph{i.e.}, $\$_{C\cdots
CC}>\$_{D\cdots DD}$, then the $n$-th agent chooses $\phi=\pi/l$ to
minimize $\sin^{2}(\phi-\frac{\pi}{l})$. As a result, $P_{C\cdots
CC}=\cos^{2}(\theta/2)$. Since $\lambda_{5}$ is satisfied,
\emph{i.e.}, $\$_{C\cdots CC}>\$_{C\cdots
CD}\cos^{2}(\pi/l)+\$_{D\cdots DC}\sin^{2}(\pi/l)$, then the $n$-th
agent prefers $\theta=0$, which leads to $P_{C\cdots CC}=1$ and let
$\$_{n}$ be its maximum $\$_{C\cdots CC}$. Therefore, the optimal
value of $(\theta_{n},\phi_{n})$ for the $n$-th agent is
$(0,\pi/l)$.

\emph{Part 2}. Let the first $l$ agents be $\hat{N}$. Consider the
following strategy profile: each agent $j=1, \cdots, l$ submits
$(\theta_{j}, \phi_{j})=(0,\pi/l)$; each agent $j=(l+1), \cdots,
(n-l+1)$ submits $(\theta_{j}, \phi_{j})=(0, 0)$, $(a_{j}, t_{j},
z_{j})=(\hat{a}, \hat{t}, 0)$, then we will prove the optimal values
of $(\theta_{n},\phi_{n})$ and $(a_{n}, t_{n}, z_{n})$ for the
$n$-th agent are $(0,0)$ and $(\hat{a}, \hat{t}, 0)$.

As shown before,
\begin{equation*}
  \hat{C}_{l}=\hat{\omega}(0,\pi/l)=\begin{bmatrix}
  e^{i\frac{\pi}{l}} & 0 \\
  0 & e^{-i\frac{\pi}{l}}
\end{bmatrix}_{2\times 2},\quad
  \hat{C}_{l}\otimes\hat{C}_{l}=\begin{bmatrix}
  e^{i\frac{2\pi}{l}} &   &   & \\
  & 1 &   & \\
  &   &1  & \\
  & &  & e^{-i\frac{2\pi}{l}}
\end{bmatrix}_{2^{2}\times 2^{2}},
\end{equation*}
\begin{equation*}
\underbrace{\hat{C}_{l}\otimes\cdots \otimes
\hat{C}_{l}}\limits_{l}=\begin{bmatrix}
  -1 &   &   & \\
  & * &   & \\
  &   &\cdots  & \\
  & &  & -1
\end{bmatrix}_{2^{l}\times 2^{l}}.
\end{equation*}
Note that
\begin{equation*}
\underbrace{\hat{I}\otimes\cdots \otimes
\hat{I}}\limits_{n-l-1}=\begin{bmatrix}
  1 &   &   & \\
  & 1 &   & \\
  &   &\cdots  & \\
  & &  & 1
\end{bmatrix}_{2^{n-l-1}\times 2^{n-l-1}},
\end{equation*}
thus,
\begin{equation*}
\underbrace{\hat{C}_{l}\otimes\cdots \otimes
\hat{C}_{l}}\limits_{l}\otimes\underbrace{\hat{I}\otimes\cdots
\otimes \hat{I}}\limits_{n-l-1}=\begin{bmatrix}
  -1 &   &   & \\
  & * &   & \\
  &   &\cdots  & \\
  & &  & -1
\end{bmatrix}_{2^{n-1}\times 2^{n-1}}.
\end{equation*}
Suppose the $n$-th agent chooses arbitrary parameters $(\theta,
\phi)$ in his strategy $(\theta, \phi, a_{n}, t_{n}, z_{n})$, let
\begin{equation*}
  \hat{\omega}_{n}(\theta, \phi)=\begin{bmatrix}
  e^{i\phi}\cos(\theta/2) & i\sin(\theta/2)\\
  i\sin(\theta/2) & e^{-i\phi}\cos(\theta/2)
\end{bmatrix},
\end{equation*}
then,
\begin{align*}
\underbrace{\hat{C}_{l}\otimes\cdots \otimes
\hat{C}_{l}}\limits_{l}&\otimes\underbrace{\hat{I}\otimes\cdots
\otimes \hat{I}}\limits_{n-l-1}\otimes\hat{\omega}_{n}(\theta,
\phi)\\
&=\begin{bmatrix}
  -e^{i\phi}\cos(\theta/2) &*   &   &   &   &    & \\
  -i\sin(\theta/2) &*   &   &   &   &    & \\
   &  & *  & *  &   &    & \\
   &  & *  & *  &   &    & \\
   &  &    &    &\cdots&    & \\
   &  &    &    & & * & -i\sin(\theta/2) \\
   &  &    &    & & * & -e^{-i\phi}\cos(\theta/2)
\end{bmatrix}_{2^{n}\times 2^{n}}.
\end{align*}
\begin{align*}
\overrightarrow{\psi}_{2}=[\underbrace{\hat{C}_{l}\otimes\cdots
\otimes
\hat{C}_{l}}\limits_{l}&\otimes\underbrace{\hat{I}\otimes\cdots
\otimes \hat{I}}\limits_{n-l-1}\otimes\hat{\omega}_{n}(\theta,
\phi)]\overrightarrow{\psi}_{1}=\frac{1}{\sqrt{2}}\begin{bmatrix}
  -e^{i\phi}\cos(\theta/2)\\
  -i\sin(\theta/2)\\
  0\\
  \cdots\\
  0\\
  \sin(\theta/2)\\
  -ie^{-i\phi}\cos(\theta/2)
\end{bmatrix}_{2^{n}\times 1},
\end{align*}
\begin{align*}
\overrightarrow{\psi}_{3}=\hat{J}^{+}\overrightarrow{\psi}_{2}=\begin{bmatrix}
  -\cos(\theta/2)\cos\phi\\
  -i\sin(\theta/2)\\
  0\\
  \cdots\\
  0\\
  \cos(\theta/2)\sin\phi
\end{bmatrix}_{2^{n}\times 1}.
\end{align*}
The probability distribution $\Delta$ is computed from
$\overrightarrow{\psi}_{3}$:
\begin{align*}
&P_{C\cdots CC}=\cos^{2}(\theta/2)[1-\sin^{2}\phi],&&P_{C\cdots CD}=\sin^{2}(\theta/2)\\
&P_{D\cdots DC}=0, &&P_{D\cdots DD}=\cos^{2}(\theta/2)\sin^{2}\phi
\end{align*}
Obviously,
\begin{equation*}
P_{C\cdots CC}+P_{C\cdots CD}+P_{D\cdots DC}+P_{D\cdots DD}=1.
\end{equation*}
Consider the payoff to the $n$-th agent,
\begin{align*}
  \$_{n}=\$_{C\cdots CC}\cos^{2}(\theta/2)[1-\sin^{2}\phi]+\$_{C\cdots CD}\sin^{2}(\theta/2)
  +\$_{D\cdots DD}\cos^{2}(\theta/2)\sin^{2}\phi
\end{align*}
Since $\lambda_{4}$ is satisfied, \emph{i.e.}, $\$_{C\cdots
CC}>\$_{D\cdots DD}$, then the $n$-th agent chooses $\phi=0$. As a
result,
\begin{align*}
  \$_{n}=\$_{C\cdots CC}\cos^{2}(\theta/2)+\$_{C\cdots CD}\sin^{2}(\theta/2)
\end{align*}
Since the $n$-th agent belongs to $N\backslash\hat{N}$, by condition
$\lambda_{3}$, $(a_{n}, t_{n}, z_{n})$ can be chosen as $(\hat{a},
\hat{t}, 0)$. According to Step 5 of \emph{MessageComputing},
$\$_{C\cdots CC}=\$_{C\cdots CD}$. Thus, $\$_{n}=\$_{C\cdots CC}$.
In this case, $\hat{\omega}_{n}(\theta,\phi)$ can be chosen as
$\hat{\omega}(0,0)=\hat{I}$.

By symmetry, in state $\bar{t}$, consider the following strategy:
each agent $j\in N\backslash\hat{N}$ submits $(\theta_{j},
\phi_{j})=(0, 0)$, $(a_{j}, t_{j}, z_{j})=(\hat{a}, \hat{t}, 0)$;
each agent $j\in\hat{N}$ submits $(\theta_{j}, \phi_{j})=(0,\pi/l)$.
Then this strategy profile is a Nash equilibrium of $\Gamma$ in
state $\bar{t}$, and the final outcome implemented in Nash
equilibrium is $\hat{a}\notin F(\bar{t})$.

Note: The proof of Lemma 1 cites the derivation of Eq. (25)
\cite{Flitney2007}.

%*****************************************************************
\end{document}